\documentclass{ws-p8-50x6-00}

\begin{document}

\title{
\vspace{-3mm}
\rightline{\small IFUP-TH 2001/36}
\vspace{8mm}
RENORMALIZATION-GROUP FLOW IN THE 3D GEORGI-GLASHOW MODEL}

\author{Dmitri Antonov}

\address{INFN-Sezione di Pisa, Universit\'a degli studi di Pisa,
Dipartimento di Fisica, Via Buonarroti, 2 - Ed. B - I-56127 Pisa, 
Italy\\
and\\
Institute of Theoretical and Experimental Physics, B. Cheremushkinskaya 
25, RU-117 218 Moscow, Russia\\
E-mail: antonov@df.unipi.it}


\maketitle

\abstracts{
The renormalization-group (RG) flow in the finite-temperature 
(2+1)-dimensional Georgi-Glashow model is explored. This is done
in the limit when the squared electric coupling constant is
much larger than the mass of the Higgs field. The novel equation 
describing the evolution of the Higgs mass is derived and 
integrated along the separatrices of the RG flow
in the limit when the original theory reduces 
to the 2D XY model. In particular, it is checked 
that in the vicinity of the phase-transition point, there exists 
a range of parameters allowing to 
the Higgs mass evolved along some of the separatrices 
to remain much smaller than 
the squared electric coupling constant.}

\section{The model}

(2+1)D Georgi-Glashow model is known to be the famous example of a 
theory allowing for an analytical description of confinement~\cite{1}.
However, the finite-temperature effects in this theory were addressed
only recently. Namely, first in ref.~\cite{2} the phase transition 
associated with the binding of monopoles into molecules has been 
studied and then in ref.~\cite{3}, there has been explored another
phase transition corresponding to the deconfinement of charged
W-bosons. In this talk, we shall concentrate ourselves at the first
of these two phase transitions, but account also for the effects
brought about by the Higgs field. In this way, we shall follow
the analysis performed in ref.~\cite{4}.

The Euclidean action of the (2+1)D Georgi-Glashow model has the 
following form

\begin{equation}
\label{GG}
S=\int d^3x\left[\frac{1}{4g^2}\left(F_{\mu\nu}^a\right)^2+
\frac12\left(D_\mu\Phi^a\right)^2+\frac{\lambda}{4}\left(
\left(\Phi^a\right)^2-\eta^2\right)^2\right],
\end{equation}
where the Higgs field $\Phi^a$ transforms by the adjoint representation, 
and $D_\mu\Phi^a\equiv\partial_\mu\Phi^a+\varepsilon^{abc}A_\mu^b
\Phi^c$.
In the one-loop approximation, the 
partition function of this theory reads~\cite{dietz}

$$
{\cal Z}=1+\sum\limits_{N=1}^{\infty}\frac{\zeta^N}{N!}
\left[
\prod\limits_{i=1}^{N}\int d^3z_i\sum\limits_{q_i=\pm 1}^{}\right]
\times
$$

\begin{equation}
\label{1}
\times\exp\left\{-\frac{g_m^2}{2}\left[\int d^3xd^3y\rho({\bf x})
D_0({\bf x}-{\bf y})\rho({\bf y})-
\sum\limits_{{a,b=1\atop a\ne b}}^{N}
D_m({\bf z}_a-{\bf z}_b)
\right]\right\}.
\end{equation}
Here, $g_m$ is the magnetic coupling constant of dimensionality 
$[{\rm length}]^{1/2}$ related to the electric one $g$
according to the equation $gg_m=4\pi$, $\rho({\bf x})=
\sum\limits_{a=1}^{N}q_a\delta\left({\bf x}-{\bf z}_a\right)$
is the density of monopole plasma with $q_a$'s standing for the 
monopole charges in the units of $g_m$. Next, in Eq.~(\ref{1}), 
$m=\eta\sqrt{2\lambda}$ is the mass of the Higgs boson and 

\begin{equation}
\label{zeta}
\zeta=\frac{m_W^{7/2}}{g}\delta\left(\frac{\lambda}{g^2}\right)
{\rm e}^{-(4\pi/g^2)m_W\epsilon\left(\lambda/g^2\right)}
\end{equation}
is the statistical weight of a single monopole (else called fugacity)
with $m_W=g\eta$ being the mass of the $W$-boson.
Here, $\epsilon$ is a slowly varying function equal to unity at the 
origin ({\it i.e.} in the Bogomolny-Prasad-Sommerfield 
limit~\cite{bps}) and $1.787\ldots$ at infinity~\cite{kirk}, whereas 
the function $\delta$ is determined by the loop corrections.
Finally, in eq.~(\ref{1}), $D_0({\bf x})\equiv1/(4\pi|{\bf x}|)$ 
is the Coulomb propagator, and 
$D_m({\bf x})\equiv{\rm e}^{-m|{\bf x}|}/(4\pi|{\bf x}|)$
is the propagator of the Higgs boson.

Notice that as it follows from eq.~(\ref{1}), in the 
Bogomolny-Prasad-Sommerfield limit, 
the interaction of two monopoles doubles for opposite and vanishes 
for equal charges. As far as the opposite limit, $m\to\infty$, is 
concerned, we apparently arrive  
there at the standard compact-QED result~\cite{1}.

The effective field theory describing the grand canonical 
partition function~(\ref{1}) can easily be obtained and 
its action reads~\cite{dietz}

\begin{equation}
\label{2}
S=\int d^3x\left[
\frac12(\nabla\chi)^2+\frac12(\nabla\psi)^2+\frac{m^2}{2}\psi^2-
2\zeta{\rm e}^{g_m\psi}\cos(g_m\chi)\right],
\end{equation}
where $\chi$ is the dual photon field, whereas the field $\psi$ is an
additional one. The latter field can be integrated out in the 
limit $g\gg\sqrt{m}$. It can be shown~\cite{4} that in this limit, 
the exponent in the last term 
on the r.h.s. of eq.~(\ref{2}) can 
be approximated by the terms not higher than the linear one.
(Note that the above inequality is implied only in the polynomial and
not in the exponential sense.) 

In such a limit, 
Gaussian integration over the field $\psi$ yields the following 
action of the dual photon field:

$$
S=\int d^3x\left[
\frac12(\nabla\chi)^2-
2\zeta\cos(g_m\chi)\right]-
$$

\begin{equation}
\label{3}
-2(g_m\zeta)^2\int d^3xd^3y\cos(g_m\chi
({\bf x}))D_m({\bf x}-{\bf y})\cos(g_m\chi({\bf y})).
\end{equation}
The last term here represents the correction to the standard
result~\cite{1}. It stems from the fact that the mass of the 
Higgs field was considered to be not infinitely large compared to the
standard Debye mass of the dual photon, $m_D=g_m\sqrt{2\zeta}$.
The respective correction to $m_D$ is positive, and the square of the 
full mass reads: $M^2=m_D^2\left(1+\frac{m_D^2}{m^2}\right)$.
Clearly, this result is valid at $m_D\ll m$ and reproduces $m_D^2$ 
in the limit $m\to\infty$.

Another relation between the dimensionful parameters in the 
model~(\ref{GG}), we shall adapt for our analysis, is 
$g\ll\eta$.
[Clearly, this inequality parallels the requirement that $\eta$ should
be large enough to ensure the spontaneous symmetry breaking from $SU(2)$
to $U(1)$.] In particular, from this relation and the
inequality $g\gg\sqrt{m}$ 
we immediately obtain:
$\frac{\lambda}{g^2}\sim\left(\frac{m}{m_W}\right)^2\ll\left(\frac{g}{\eta}
\right)^2\ll 1$. 
This means that we are working in the regime of the Georgi-Glashow 
model close to the Bogomolny-Prasad-Sommerfield limit.

Note further that in the limit $g\gg\sqrt{m}$, 
the dilute gas approximation holds perfectly for monopole plasma. 
Indeed, this approximation implies
that the mean distance between monopoles, equal to $\zeta^{-1/3}$, 
should be much larger than the inverse mass of the $W$-boson. By virtue of 
eq.~(\ref{zeta}) and the fact that the function $\epsilon$ is of the order
of unity, we obtain that this 
requirement is equivalent to the following one: 

\begin{equation}
\label{extra}
\sqrt{\frac{\eta}{g}}\delta\left(\frac{\lambda}{g^2}\right)
{\rm e}^{-4\pi\eta/g}\ll 1.
\end{equation}
Although at $\lambda\ll g^2$ 
the function $\delta$ grows, the speed of this growth 
is so that at $g\ll\eta$,
the l.h.s. of the inequality~(\ref{extra}) remains exponentially 
small~\cite{ks}.
Another consequence of this fact is that in the regime
of the Georgi-Glashow model under discussion, the Debye mass 
of the dual photon, $m_D$, remains exponentially small as well.
In particular, the inequality $m_D\ll m$, under which the full mass $M$
was derived, holds due to this smallness. Also, due to the same reason,
the mean field approximation, under which the effective field
theory~(\ref{2}) is applicable, remains valid as well with the 
exponential accuracy. This approximation means that one can 
disregard the fluctuations of individual monopoles only provided that
in the Debye volume $m_D^{-3}$ there contained a lot of them. 
This condition can formally be written as

$$\left[{\rm average}~ {\rm density}~ = \frac{\partial
\ln{\cal Z}}{V\partial\ln\zeta}\simeq 2\zeta\right]\times m_D^{-3}
\gg 1,$$
where $V$ is the 3D volume of observation. This yields the 
inequality $g^3\gg\zeta$,
which is really satisfied owing to the above-discussed exponential 
smallness of fugacity $\zeta$.

\section{The RG flow}

At finite temperature $T\equiv 1/\beta$, one should supply all the 
fields in the model~(\ref{GG}) with the periodic boundary conditions
in the time direction, with the period equal to $\beta$.
The lines of the magnetic field of a monopole thus cannot cross the 
boundary of the one-period region in the time direction and should 
go parallel to this boundary at the distances larger than $\beta$.
Therefore, monopoles separated by such distances interact via the  
2D Coulomb potential, rather than the 3D one. Recalling that the 
average distance between monopoles is equal to $\zeta^{-1/3}$, we 
conclude that at $T\ge\zeta^{1/3}$, the monopole ensemble becomes 
two-dimensional.

This result can also be obtained formally by computing the following 
sum over Matsubara frequencies:

\begin{equation}
\label{coul}
\frac{1}{|{\bf x}|}\equiv\sum\limits_{n=-\infty}^{+\infty}
\frac{1}{\sqrt{\vec x^2+(\beta n)^2}}=2T\sum\limits_{n=-\infty}^{+\infty}
K_0(2\pi T|\vec x{\,}|n)\simeq-2T\ln(\mu|\vec x{\,}|).
\end{equation}
Here, $\mu$ denotes the IR momentum cutoff, 
$\vec x\equiv\left(x^1,x^2\right)$, and 
without the loss of generality we have considered the case 
$x_0=0$. Next, $K_0$ stands for the 
modified Bessel function, which is rapidly decreasing. Owing to this fact,
the term with $n=0$ (the so-called zero Matsubara mode) dominates in the 
whole sum, 
which yields the last equality. In the same way, we obtain for the 
Yukawa propagator:

$$
\frac{{\rm e}^{-m|{\bf x}|}}{|{\bf x}|}\equiv\sum
\limits_{n=-\infty}^{+\infty}\frac{{\rm e}^{-m\sqrt{\vec x^2+
(\beta n)^2}}}{\sqrt{\vec x^2+(\beta n)^2}}=$$

\begin{equation}
\label{yuk}
=2T\sum\limits_{n=-\infty}^{+\infty}K_0\left(m|\vec x{\,}|
\sqrt{1+\left(\frac{2\pi Tn}{m}\right)^2}\right)\simeq
2TK_0(m|\vec x{\,}|).
\end{equation}
These equations mean that the strength of the 
monopole-antimonopole interaction, stemming from eq.~(\ref{1})
at finite temperature, is proportional itself to the temperature.
Owing to this fact, at low temperatures the interaction is weak,
{\it i.e.} monopoles exist in the plasma phase, whereas at the 
temperatures higher than some critical one, $T_c$, they form 
monopole-antimonopole molecules. This situation is reversed with 
respect to the standard Berezinskii-Kosterlitz-Thouless (BKT) phase 
transition in the 2D XY model~\cite{BKT}. There, the strength of the 
interaction is $T$-independent, which leads to the molecular
phase at low temperatures and to the plasma phase at high temperatures.
In our model, the critical temperature of the phase transition is then the 
one, below which the mean squared separation of a monopole and an
antimonopole in the molecule diverges. According to the 
formulae~(\ref{coul}) and~(\ref{yuk}), this separation reads

$$
\left<L^2\right>\propto\int d^2x|\vec x{\,}|^{2-\frac{8\pi T}{g^2}}
\exp\left[\frac{4\pi T}{g^2}K_0(m|\vec x{\,}|)\right].$$
This yields $T_c=g^2/(2\pi)$, which coincides with the result
obtained in ref.~\cite{2} without accounting for the Higgs field.
Clearly, $T_c\gg\zeta^{1/3}$, which means that there exists 
a broad range of temperatures where the monopole ensemble (plasma) is 
two-dimensional.

Let us now proceed with the formal RG analysis of the leading
$(m_D/m)$-part of the action~(\ref{3}), which has been performed
in ref.~\cite{4}. This part of the action can be written as

$$
S=-\int d^3x\times
$$

\begin{equation}
\label{act}
\times\left[\frac12\left(\partial_x^2+\partial_y^2
+a_1\partial_t^2\right)\chi+2\zeta a_2\cos(g_m\chi)+\left(\frac{g_m\zeta 
a_3}{m}\right)^2\cos(2g_m\chi)\right],
\end{equation}
where $a_1=a_2=a_3=1$.
The idea of derivation of the RG equations~\cite{SY} is to split the 
cutoff field into two pieces,
$\chi_\Lambda=\chi_{\Lambda'}+h$. Here, the field $h$ includes 
the modes with the momenta lying in the range between $\Lambda'$
and $\Lambda$ and all possible Matsubara frequencies. After that,
the field $h$ should be integrated out, which yields the same action,
but with another values of parameters $a_i$'s. The RG equations
can then be derived by performing the infinitesimal transformation
$\Lambda'=\Lambda-\delta\Lambda$ and comparing the deviation of the 
values of $a_i$'s from unity. In this way, we arrive at the following 
system of RG equations (see ref.~\cite{4} for some details of 
the derivation):

$$
dx=-x^3z^2dt,~~ 
dz^2=-2z^2\left(\pi x\frac{\tau}{2}\coth\frac{\tau}{2}-2\right)dt,
$$

$$
d\ln u=\left(\frac34-\pi x\frac{\tau}{2}\coth\frac{\tau}{2}\right)dt,~~
d\ln\tau=-2dt.
$$
Here, $t=\ln(T/\Lambda)$, $x=Tg_m^2/(4\pi^2)$, $u=\Lambda^{-3/4}
\sqrt{g_m\zeta m^{-1}}$, $\tau=A\Lambda/T$, $z=B\zeta/(\Lambda^2T)$
with $A$ and $B$ standing for some inessential constants.

The above-presented equation for the dimensionless $m$-dependent parameter 
$u$ is a new one with respect to the other equations. Those can be derived
independently~\cite{SY} in the limit when the Debye mass of the dual 
photon is considered to be negligibly small with respect to the Higgs one.
In the limit $t\to\infty$ (or, equivalently, $\tau\to 0$) the RG flow
for the quantities $x$ and $z$ becomes that of the 2D XY model with the 
$T$-dependent strength of the interaction.
In particular, we straightforwardly obtain the 
BKT phase transition point $z_c=0$, 
$x_c=2/\pi$, which reproduces the critical temperature $T_c$, derived 
above heuristically. In the vicinity of the BKT transition point, 
the RG trajectories stemming from the integration of the equations
for $x$ and $z$ are typical hyperbolae of the XY model. They are defined
by the equation $(x-x_c)^2-(2/\pi)^4z^2={\,}{\rm const}$.

In the same XY-model limit, the RG equation for $u$ becomes 
remarkably simple, $d\ln u=\left(\frac34-\pi x\right)dt$, 
and can be integrated along the separatrices $T=T_c$, which yields

$$\ln u=-\frac{5}{2\pi(x-x_c)}+\frac74\ln\frac{2x}{\pi|x-x_c|}
+\frac{3}{2\pi^2x^2}-\frac{1}{\pi x}+{\,}{\rm const}.$$
This equation means that at $x\to x_c+0$, $u$ exponentially vanishes
along the respective (right) separatrix, whereas at $x\to x_c-0$,
$u$ exponentially grows along the left separatrix. By 
virtue of eq.~(\ref{zeta}) 
with $\epsilon\sim 1$, the first of these facts can formally be written as

$$\frac{\eta}{g}\delta^2\left(\frac{\lambda}{g^2}\right)
{\rm e}^{-8\pi\eta/g}\ll\left(\frac{m}{g^2}\right)^2
\left(\frac{\Lambda}{\eta^2}\right)^3,$$
and the second fact corresponds to the change of the symbol "$\ll$"
to the opposite one. Clearly, varying two large parameters $(\eta/g)$
and $(g^2/m)$, as well as the ratio $(\Lambda/\eta^2)$,
we may occur in the region of validity of any of these 
two inequalities. In another words, independently along which of the 
two separatrices the theory approaches its critical point, we can adjust
the parameters $(\eta/g)$ and $(\Lambda/\eta^2)$ in such a way that
the respective inequality holds at $g\gg\sqrt{m}$. Thus, we conclude that
the theory~(\ref{act}) reduced by the RG flow to the 2D
XY model stays within the original approximation $g\gg\sqrt{m}$
when the flow drives this theory 
in the vicinity of the critical point along the separatrices 
$T=T_c$.

\section*{Acknowledgments}
The author is grateful to Prof. A. Di Giacomo for useful discussions
and to Dr. N. Agasian, in collaboration with whom the paper~\cite{4}
has been written. This work has been supported by
INFN and partially by the INTAS grant Open Call 2000, project No. 110.
And last but not least, the author acknowledges the organizers
of the Tenth Lomonosov Conference on Elementary Particle Physics
(Moscow, 23-29 August, 2001)
for an opportunity to present the above-discussed results in a very 
stimulating atmosphere.

\end{document}